\documentclass[twocolumn,showpacs,preprintnumbers,amsmath,amssymb]{revtex4}

\usepackage{graphicx,dcolumn,bm,color}
\usepackage{epsfig,subfigure,amsmath}

\begin{document}
\preprint{DRAFT FOR REVIEW}
\title{Launching and Controlling Gaussian Beams from Point Sources \\ via Planar Transformation Media}


\author{Hayrettin Odabasi$^1$}
 \altaffiliation{Corresponding author. Email: hodabasi@ogu.edu.tr}
 \author{Kamalesh Sainath$^2$}
\author{Fernando L. Teixeira$^3$}
\affiliation{$^1$Department of Electrical and Electronics Engineering, Eskisehir Osmangazi University, Eskisehir 26480, Turkey,}%
\affiliation{$^2$Sandia National Laboratories, Albuquerque, New Mexico 87185, USA,}%
\affiliation{$^3$ElectroScience Laboratory, The Ohio State University, Columbus Ohio 43212, USA.}%

\date{\today}


\begin{abstract}

Based on operations prescribed under the paradigm of Complex Transformation Optics (CTO)~\cite{Teixeira_to_99,Teixeira_to_00,Odabasi_to_11,Popa_to_11}, it was recently shown in~\cite{Castaldi_to_13} that a complex source point (CSP) can be mimicked by a parity-time ($\mathcal{PT}$) transformation media. Such coordinate transformation has a mirror symmetry for the imaginary part, and results in a balanced loss/gain metamaterial slab. A CSP produces a Gaussian beam and, consequently, a point source placed at the center of such metamaterial slab produces a Gaussian beam propagating away from the slab. Here, we extend the CTO analysis to non-symmetric complex coordinate transformations as put forth in ~\cite{Savoia_to_16} and verify that, by using simply a (homogeneous) doubly anisotropic gain-media metamaterial slab, one can still mimic
a CSP and produce Gaussian beam. In addition, we show that a Gaussian-like beams can be produced by point sources placed {\it outside} the slab as well~\cite{Savoia_to_16}. By making use of the extra degrees  of freedom (real and imaginary part of the coordinate transformation) provided by CTO, the near-zero requirement on the real part of the resulting constitutive parameters can be relaxed to facilitate potential realization of Gaussian-like beams. We illustrate how beam properties such as peak amplitude and waist location can be controlled by a proper choice of (complex-valued) CTO Jacobian elements. In particular, the beam waist location may be moved bidirectionally by allowing for negative entries in the Jacobian (equivalent to inducing negative refraction effects). These results are then interpreted in light of the ensuing CSP location.

\end{abstract}


\maketitle


\section{Introduction}
 There has been an increased interest in complex transformation optics (CTO)~\cite{Teixeira_to_99,Teixeira_to_00,Odabasi_to_11,Popa_to_11,Castaldi_to_13,Savoia_to_16} as a tool for designing
metamaterials with functionalities beyond the reach of conventional TO~\cite{Pendry_to_06,Leonhardt_to_06a,Leonhardt_book}. Both TO and CTO rely on the known duality between metric and material properties in Maxwell's equations~\cite{Teixeira_to_99,Teixeira_to_00} to provide a pathway for designing metamaterial blueprints that mimic a change on the metric of space~\cite{Schurig_mm_cloak_06}. While TO is restricted to real-valued coordinate transformations, CTO employs complex-valued space transformations in a frequency-domain representation. Noteworthy examples of CTO-derived media are perfectly matched layers (PMLs), which are reflectionless absorbers extensively used to truncate the spatial domain in computational simulations~\cite{Berenger_pml_94,Chew_pml_94,Sacks_pml_95}. Although their introduction was initially motivated strictly by simulation needs, PMLs also serve as blueprints for anisotropic absorbers~\cite{Sacks_pml_95,Teixeira_pml_97,Ziolkowski_pml_97,Teixeira_pml_03,Odabasi_to_11,Sainath_pml_15}.

It is well known that a Gaussian beam can be well approximated in the paraxial region as the field produced  by a complex source point (CSP)~\cite{Keller_csp_71,Deschamps_csp_71,Felsen_csp_76,Heyman_csp_01,Tap_csp_07}. The CSP field is an exact solution of Maxwell's equation that can be derived via analytic continuation to complex-valued coordinates. Recently, it was shown in~\cite{Castaldi_to_13} that a CSP fields can be blue obtained via parity-time ($\mathcal{PT}$) metamaterials with a particular mirror-symmetric CTO transformation. Such CTO profile results in a balanced gain/loss transformation media ($\mathcal{PT}$ metamaterial slab). Due to the mirror-symmetric nature of that transformation, in order to produce the CSP field, the source needs to be placed inside the metamaterial slab. More recently, CTO was extended to manipulate CSP-based wave objects~\cite{Savoia_to_16} via non-Hermitian transformation slabs. With the source is placed outside of the slab, the study in~\cite{Savoia_to_16} has shown how fields on one side of the slab can be interpreted as generated from an `image' CSP on the opposite side. 

In this work, we further develop on~\cite{Castaldi_to_13} and ~\cite{Savoia_to_16} to study how CSP fields can be generated and their properties controlled by impedance-matched slabs comprised of doubly anisotropic gain-media metamaterials without $\mathcal{PT}$ symmetry. Notably, we show how beam properties such as peak amplitude and waist (focus) location can be modified by a proper choice of (complex-valued) elements of the Jacobian of the CTO transformation. In particular, control of the beam waist location may be extended by allowing for negative entries in the Jacobian, which is equivalent to inducing negative refraction effects. We also interpret these results in light of the ensuing (effective) CSP location. To simplify the discussion, we employ the terms CSP field and Gaussian beam interchangeably in this paper; however, it should kept in mind that the equivalence of the CSP field to a Gaussian beam is only approximate and restricted to the paraxial region.

%
%
%
\section{Formulation}
Throughout the paper the $e^{-i\omega t}$ convention is assumed and omitted. Let us assume the following CTO mapping from real space $\left(x,y,z\right)$ to complex space $\left(x',y,z\right)$:
\begin{equation}\label{to_xyz}
x'(x)=
\left\{
\begin{array}{lr} 
x'(-d/2) + d/2 + x  \hspace{0.6cm} \text{if} \hspace{0.3cm} x \le -d/2 
\vspace{0.15cm} \\ 
d/2 + \int^{x}_{d/2} {s_x\left(x\right)} {dx} \hspace{0.7cm} \text{if} \hspace{0.3cm} -d/2 \leq x \leq d/2 
\vspace{0.15cm} \\
x \hspace{3.4cm} \text{if} \hspace{.3cm} d/2 \leq x 
\end{array}
\right.
\end{equation}
\noindent where $s_x(x) = a_x(x) + i \sigma_x(x)$ is a complex stretching factor, and $d$ is the slab thickness.
Using the (C)TO approach, the associated constitutive tensors~\footnote{Note that our definitions of flat-space and
“deformed” (“complex-space”) coordinates follow much of the
PML literature, and are reversed w. r.  t. the TO literature. The final material properties are,
of course, independent of coordinate convention.} are obtained as $\left[\epsilon\right] = \epsilon_0 [\Lambda]$ and $\left[\mu\right] = \mu_0 [\Lambda]$,
with $[\Lambda] = \text{det}(\left[S\right])^{-1} \left[S\right] \left[S\right]^T$, where $[S]$ is the Jacobian $(\partial x, \partial y, \partial z)/(\partial x', \partial y', \partial z')$ of the transformation in
eq.~(\ref{to_xyz})~\cite{Pendry_to_06,Leonhardt_book,Teixeira_to_99}, i.e.,
\begin{equation} \label{Lambda}
\left[\Lambda\right]=
\left\{
\begin{array}{lr} 
\text{diag}(1,1,1) \hspace{1.25cm} \text{if} \hspace{0.3cm}  x \le -d/2 
\vspace{0.1cm} \\ 
\text{diag}(s_x^{-1},s_x,s_x) \hspace{0.55cm} \text{if} \hspace{0.3cm}   -d/2 \leq x \leq d/2   
\vspace{0.1cm} \\
\text{diag}(1,1,1) \hspace{1.25cm} \text{if} \hspace{0.3cm}  d/2 \leq x 
\end{array}
\right.
\end{equation}
 Note that the coordinate transformation needs to be continuous to avoid reflections or scattering. Clearly, this is satisfied by Eq.(\ref{to_xyz}) as long as $a_x$ and $\sigma_x$ are bounded functions. In this case, the resulting metamaterial slab is impedance-matched to free-space for all incidence angles~\cite{Sacks_pml_95,Teixeira_pml_97,Ziolkowski_pml_97,Teixeira_pml_03}.
 The material tensor corresponds to a lossy or gain media if $\sigma_x$ is chosen positive or negative, respectively~\footnote{Note that the integral in Eq.~(\ref{to_xyz}) is taking along {\it decreasing} values of $x$ so that negative $\sigma_x$ implies
 $\Im m [x']>0$  in Eq.~(\ref{to_xyz}) and hence a gain medium.}. The parameter  $a_x$ controls the real stretching and can be chosen so as to increase (if $a_x>1$) or decrease ($a_x<1$) the {\it electric} size of the slab. In particular, the electric size of the slab shrinks to zero in the limit
$a_x \rightarrow 0$, which means that no phase accumulation occurs as the wave propagates though the slab. In addition, the choice $a_x < 0$ produces negative refraction effects~\footnote{These cases resemble, but are not identical to, (isotropic) zero-index and double-negative media, respectively, because we deal here with (doubly) anisotropic media~\cite{Kuzuoglu_to_06}.}.

 Here we compare two alternative choices for the $\sigma_x$ parameter:
\begin{enumerate}
\item[(a)] {\it Balanced loss/gain} $\mathcal{PT}$ {\it media}~\cite{Castaldi_to_13}:
\begin{equation} \label{tr1}
\sigma_x=\mp 2b/d \hspace{0.4cm} \text{for}  \hspace{0.2cm} x \gtrless 0, \, \, |x| \leq d/2, 
\end{equation} 
with $b>0$. Due to the mirror symmetry of this $\sigma_x$ profile, the material tensors are $\mathcal{PT}$ symmetric, i.e. $\epsilon\left(x\right)=\epsilon^*\left(-x\right)$ and the metamaterial slab corresponds to a balanced loss/gain media. The corresponding profile of $\Im m [x']$ is shown the the red curve in Fig. \ref{to_xyz}.
\item[(b)]  {\it Doubly anisotropic gain-media}:
\begin{equation} \label{tr2}
\sigma_x=-b/d \hspace{0.4cm} \text{for} \hspace{0.2cm} |x|\leq d/2, 
\end{equation} 
again with $b>0$. The corresponding behavior of $\Im m [x']$ is shown by the green curve in Fig. \ref{to_xyz}.
\end{enumerate}
By choosing $a_x=0$ in both cases, one obtains $\Re e[x']$ as shown by the blue curve in Fig. \ref{to_xyz}.
Both cases provide the necessary imaginary component required for the realization of CSP. 
However, there is a major difference between them in that for $x<-d/2$ (i.e. outside the slab)
 $\Im m[x'] =0$ in case (a) whereas $\Im m[x'] >0$ in case (b). Note that this occurs even though $\left[\Lambda\right]= \text{diag}(1,1,1)$ (corresponding to  free-space) in that region for both cases. At first sight, this result may seem paradoxical, but really what happens in this case is that the doubly anisotropic gain-medium metamaterial slab amplifies the fields from sources placed {\it outside} it, so that in the transformed problem the point source appears mapped to complex space. This enables such metamaterial slab to be used as a Gaussian beam launcher. Note that such coordinate transformation is bidirectional; in other words, the same effect is obtained whether we place the source at $x \leq -d/2$ or $x \geq d/2$. Here, we placed the source at $x \leq -d/2$ to illustrate that the source appears to reside on a complex position (Fig.~1).

\begin{figure}[t]
\centering
\includegraphics[width=0.49\textwidth]{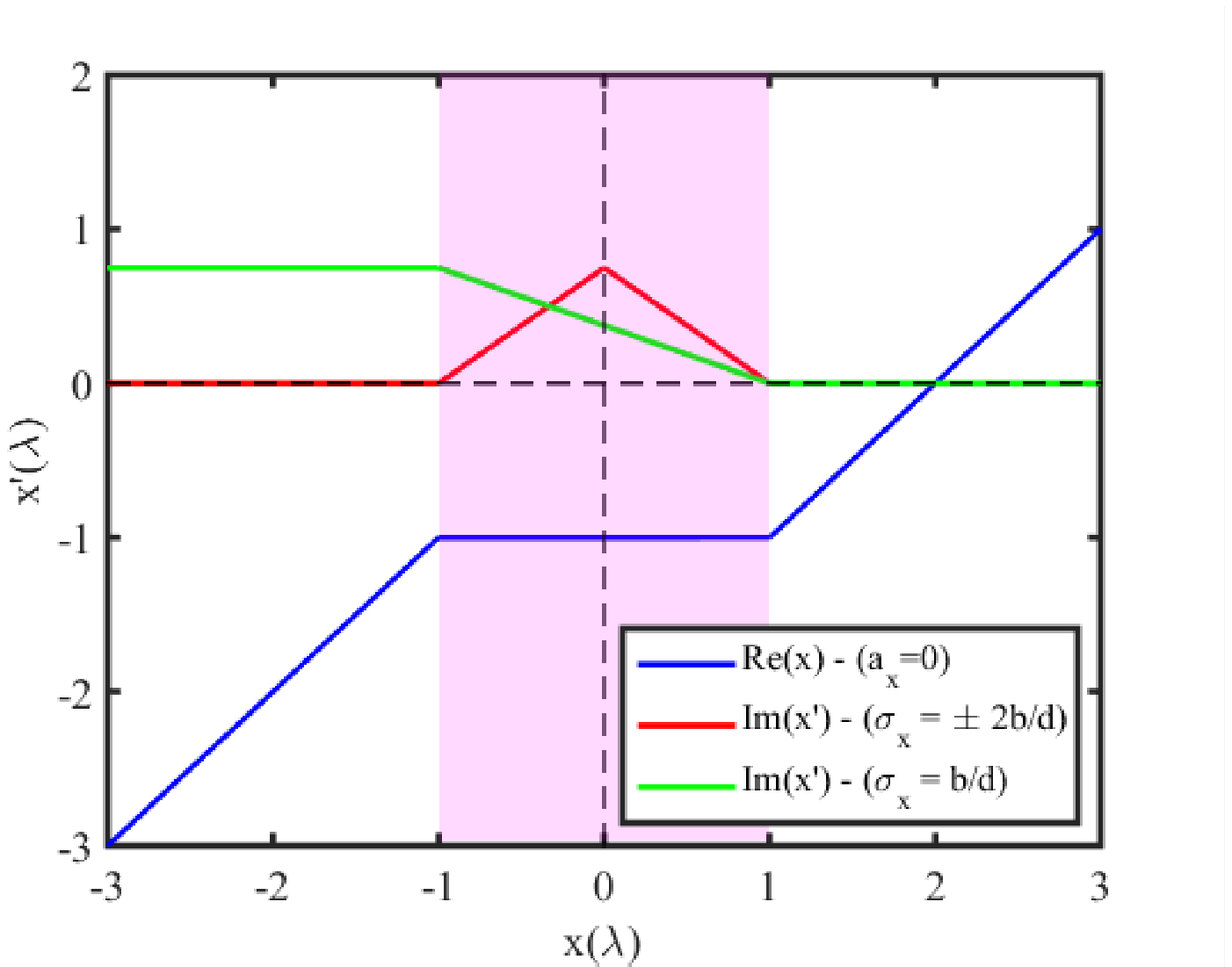}
\caption{Coordinate transformations showing the real and imaginary parts of the transformed coordinate $x'$ as a function of the original coordinate $x$, as entailed by Eqs.~(\ref{to_xyz}), (\ref{tr1}), and (\ref{tr2}).
Colored area indicates the metamaterial slab region, $|x| \leq d/2$ with $d=2\lambda$. 
The red curve corresponds to $\Im m[x']$ with $\sigma_x=\mp 2b/d$ for $x \gtrless 0$.
The green curve corresponds to $\Im m[x']$ with $\sigma_x=-b/d$.
The blue curve shows $\Re e[x']$, from $a_x=0$ for $|x| \leq d/2$. Note that with $\sigma_x=\mp 2b/d$ , $\Im m[x']$ reverts to zero outside of the slab. On the other hand, with $\sigma_x=-b/d$  there is a remnant positive value for $\Im m[x']$ in the transformed space outside the slab. The value of $b$ is chosen as $0.75 \lambda$.}
\label{fig1}
\end{figure}

\begin{figure*}[t]
\centering
\includegraphics[width=0.99\textwidth]{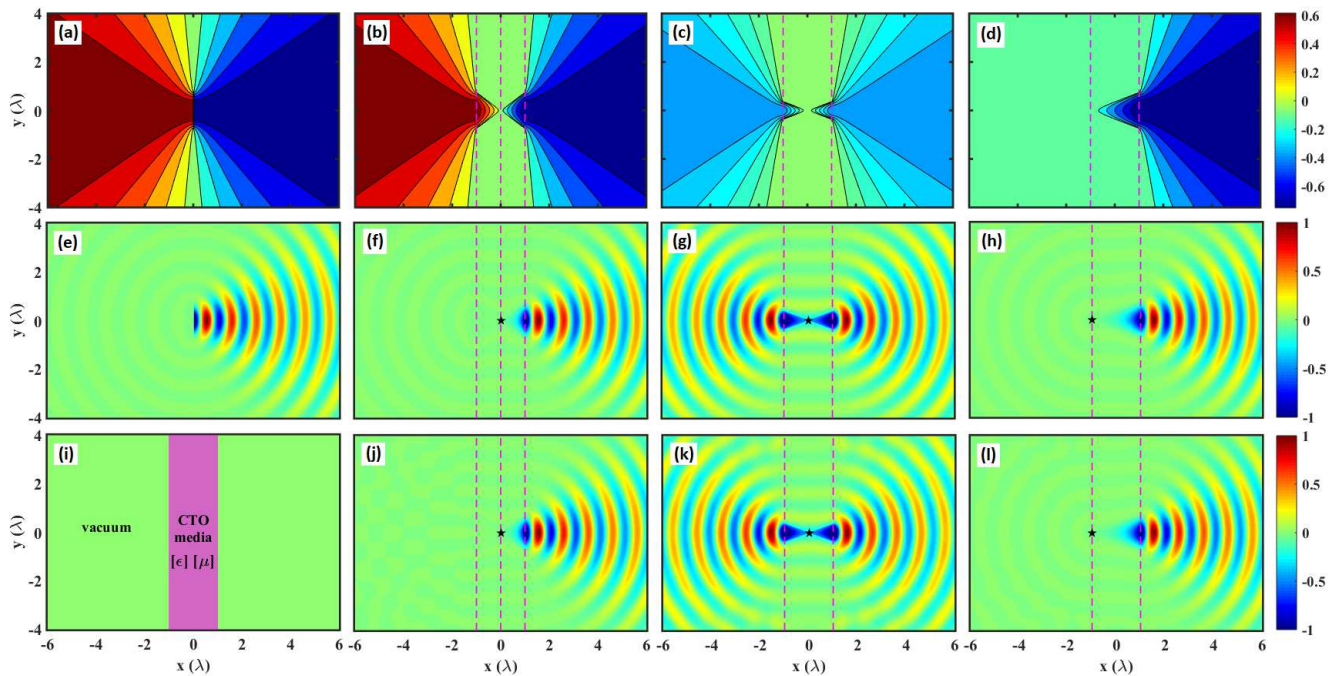}
\caption{Summary of CTO approach to obtain blueprints for planar metamaterial slabs that generate Gaussian beams from a point source placed in or next to it. (a): Contour plots of $\Im m[R']$ for a CSP placed at $(ib,0)$ in free-space, with $b=0.75 \lambda$. (b): Contour plots of $\Im m[R']$ due to point source placed at the center of the $PT$-symmetric metamaterial slab. Here, the slabs comprise the region  $|x|\leq d/2$ with $d=2\lambda$. (c, d): Contour plots of $\Im m[R']$ for a point source located at the center and at the left boundary, respectively, of a doubly anisotropic gain-medium metamaterial slab with $a_x=0$ and $\sigma_x=-b/d$. (e--h): Field distributions of $\Re e [E_z]$ based on the analytical solution, Eqs. (\ref{Ez1}) and (\ref{R}), corresponding to cases (a-d), respectively. The source points are indicated by the star symbols. (i) Geometry of the problem. The point source is placed inside or next to the metamaterial slab. (j-l): FE simulation results with metamaterial slabs corresponding to (b-d), respectively. In all field plots of this paper, $\Re e [E_z]$ is normalized to the $[-1,1]$ interval. Note that in order to avoid branch cut singularities we have chosen $a_x=0.001$.}
\label{fig2}
\end{figure*}

\begin{figure}[htbp]
\centering
\includegraphics[width=0.45\textwidth]{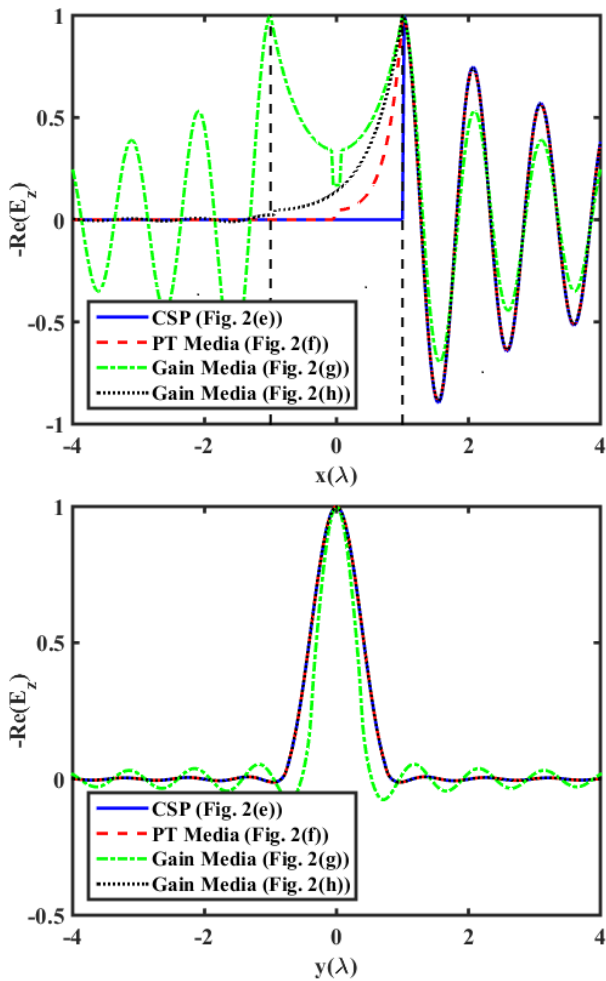}
\caption{(a, b): Horizontal ($y=0$) and vertical ($x=\lambda+\lambda/100$) cut plots of $\Re e[E_z]$ for the cases in Fig.~2(e,f,g,h).}
\label{fig3}
\end{figure}

\begin{figure*}[t]
\centering
\includegraphics[width=0.95\textwidth]{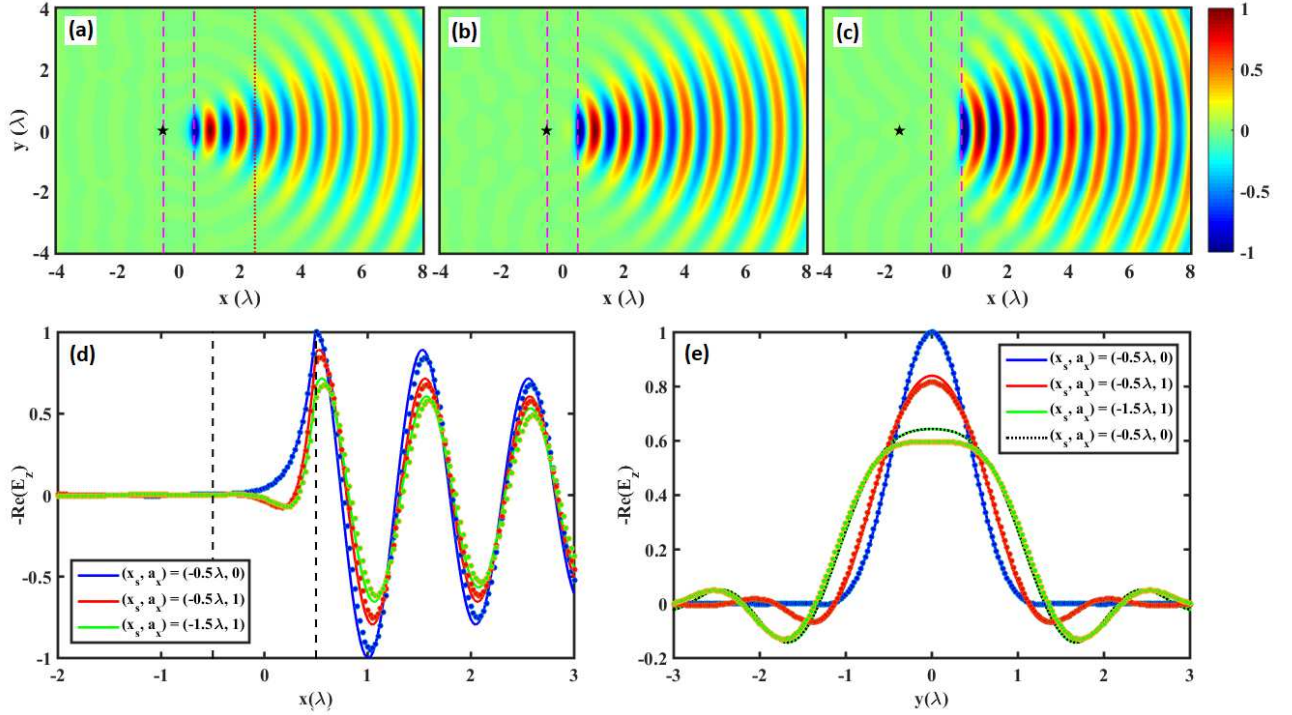}
\caption{Gaussian-like beam produced from point sources, as indicated by the star symbols, at different locations $x_s$ next to doubly anisotropic gain-media slabs with different values for the real stretching parameter $a_x$. In all cases, $d=\lambda$ and $\sigma_x=-2b/d$. (a-c): FE simulation results for three different ($(x_s,a_x)$) choices, namely, $(x_s, a_x)=(-0.5\lambda,0)$, $(x_s, a_x)=(-0.5\lambda,1)$, and $(x_s, a_x)=(-1.5\lambda,1)$ respectively. (d): $\Re e[E_z]$  along the horizontal cut $y=0$ for various choices of $(x_s, a_x)$, as indicated. (e): $\Re e[E_z]$ along the vertical cuts $x=0$ for the above three cases together with a vertical cut of
$\Re e[E_z]$ at $x=2.5\lambda$ (red dashed lines) for $(x_s, a_x)=(-0.5\lambda,0)$. The point source locations are indicated by the star symbols. Note that for $a_x=0$,  as shown in (a) and indicated by the blue curve in (d), there is no phase progression inside the slab, only amplification. In contrast, for $a_x=1$ there is a phase progression inside the slab. As a consequence, the $\Re e[E_z]$ distribution along the vertical cuts $x=2.5\lambda$  in (a) and $x=0$ in (e) are equal to each other, compare also the green and dashed lines in (e). For cut plots, the solid lines represent the analytical solution and the dots represents the numerical simulation results. All figures are normalized based on Fig.~4(a) case.}
\label{fig4}
\end{figure*}

\begin{figure*}[htbp]
\centering
\includegraphics[width=0.95\textwidth]{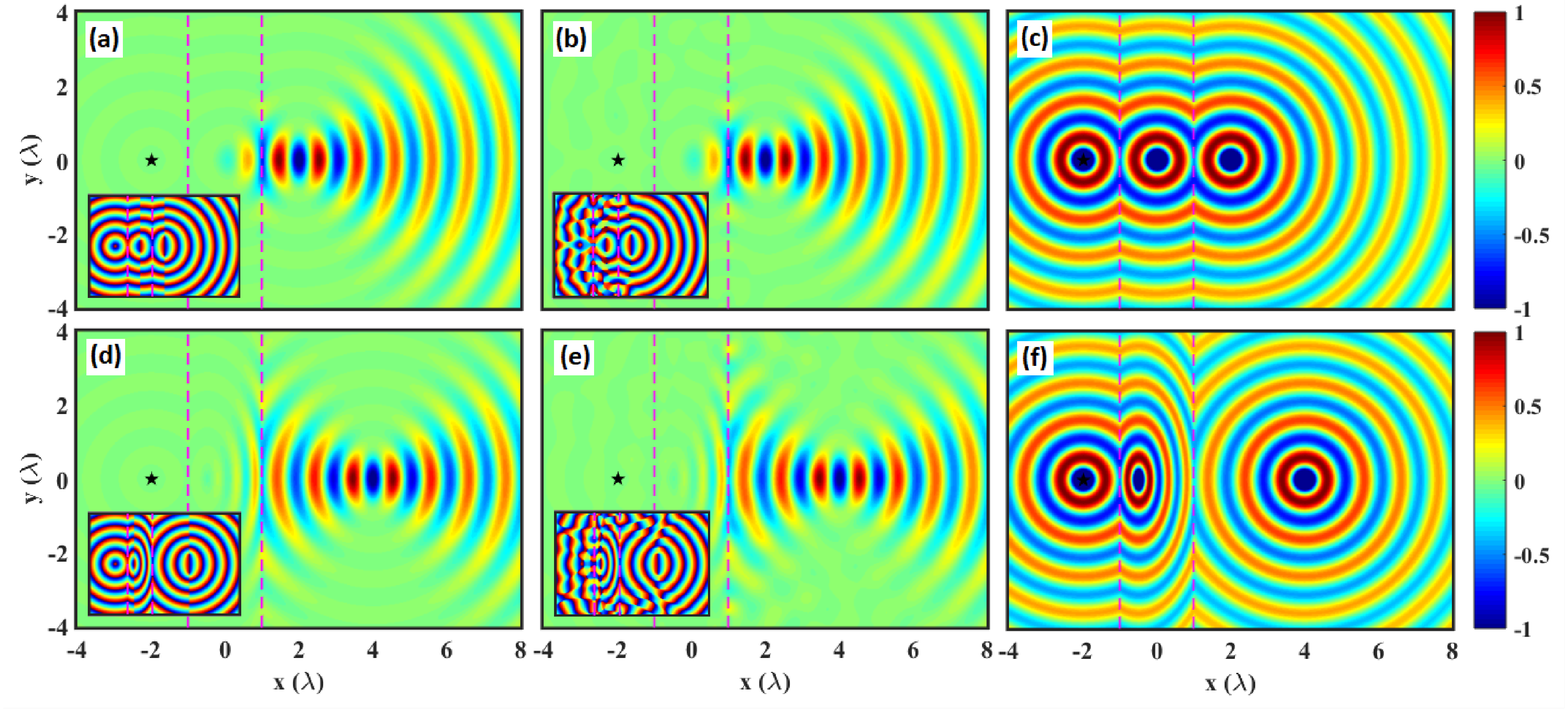}
\caption{Effect of negative $a_x$ on the Gaussian beam generation. (a,d): $\Re e [E_z]$ distributions based on the analytic solution, Eqs. (\ref{Ez1}) and (\ref{R}), for a CSP mapped by $a_x = -1$ and $a_x =-2$ respectively, and with
$\sigma_x=-b/d$ in Eq.~(\ref{to_xyz}). (b, e): Corresponding $\Re e [E_z]$ field distributions based on FE simulations with doubly anisotropic gain-media metamaterial slabs. The insets show the phase distributions to highlight the foci. (c, g): Corresponding analytical solutions when $\sigma_x=0$ (Veselago slab).}
\label{fig5}
\end{figure*}

\section{Results and Discussion}

The field solution in the transformed coordinates can be found via analytic continuation of the known Green's function. In the transformed coordinates, the field due to a point source in two dimensions (i.e. a line source in three dimensions) given by $\vec{J'}=\hat{z} \delta\left(x'-x'_s\right) \delta\left(y'-y'_s\right)$ is obtained analytically as~\cite{Harrington_book}:
\begin{equation} \label{Ez1}
\vec{E}' = -{\hat z}\frac{I_0 k_0 \eta_0}{4} H_0^{(1)}\left(k_\rho R'\right)  
\end{equation}
\noindent where $H_0^{(1)}$ is the Hankel function of first kind and zeroth order. The solution in the transformed coordinates is found by substituting the complex distance $R'$ given by
\begin{equation} \label{R}
R' = \sqrt{(x'-x'_s)^2+(y'-y'_s)^2}  
\end{equation}

\noindent where $(x'_s,y'_s)$ is the CSP location associated to a point source at $(x_s,y_s)$ after the transformation given by Eq.~\ref{to_xyz}. In order to obtain the proper field solution, a branch cut with $\Re e[R']>0$ (the so-called source-type solution) is chosen~\cite{Keller_csp_71,Deschamps_csp_71,Felsen_csp_76,Heyman_csp_01,Tap_csp_07}. The actual fields ${\vec E}$ and sources ${\vec J}$ in real space are found simply~\cite{Pendry_to_06,Leonhardt_book,Teixeira_to_99} by
\begin{subequations}
\begin{equation} \label{Ez2}
{\vec E} = \left[S^{-1}\right]^T \cdot {\vec E}'.  
\end{equation}
\begin{equation}
{\vec J} = \text{det}(\left[S\right])^{-1} \left[S\right] \cdot {\vec J}'.  
\end{equation}
\end{subequations}
In what follows, we show analytical results for the CSPs produced by various CTO mappings together with simulation results based on the Comsol$^\text{TM}$ finite element (FE) software
for point sources placed near or inside metamaterial slabs with constitutive tensors given by Eq.~(\ref{Lambda}-\ref{tr2}). In the following examples, we assume (unless otherwise stated) $b=0.75 \lambda$ and $d=2 \lambda$, where $\lambda$ is the free-space wavelength.

\subsection{Results for $a_x=0$}
Because the CSP field behavior is determined by both the real and imaginary part of $R'$, one can interpret the field behavior by examining the profile of $R'$ once the branch cut is specified. Fig.~2~(a,e) shows contour plots of $\Im m[R']$ and $\Re e[E_z]$ based on the analytical solution for a standard CSP at $(x_s,y_s)=(ib,0)$, where the Gaussian beam distribution is clearly visible~\cite{Keller_csp_71,Deschamps_csp_71,Felsen_csp_76,Heyman_csp_01,Tap_csp_07}. Fig.~2~(b,f) shows $\Im m[R']$ and $\Re e[E_z]$ for a point source placed at $(x_s,y_s)=(0,0)$ followed by transformation set by Eq.~(\ref{tr1}) so that $(x'_s,y'_s)=(ib,0)$, corresponding to the balanced loss/gain $\mathcal{PT}$ case. The associated constitutive tensors have $\left[\Lambda\right]= \mp \text{diag}(id/2b,-i2b/d,-i2b/d)$ for $x \gtrless 0$ and $\left|x\right| \leq d/2$~\footnote{Similar results can be obtained with other mirror-symmetric functions such as, for example, $\sigma_x=\sin(\pi 2x/d)$; however, in this case the transformation media becomes inhomogeneous since the Jacobian is an $x$-dependent function.}. Note that in this case, there is no Gaussian beam generation if the point source is placed at $|x| \geq d/2$ (i.e. outside the slab) since the corresponding CSP would revert to a real-valued point. This can also be understood by the fact, for a balanced loss-gain media, any field amplification over the gain section of the slab would then be compensated by field attenuation over the loss section or vice versa. Fig.~2~(c,g) shows the contour plots of $\Im m[R']$ and $\Re e[E_z]$ for a point source placed at $(x_s,y_s)=0$ followed by the transformation defined in Eq.~(\ref{tr2}) so that $(x'_s,y'_s)=(ib/2,0)$. This corresponds to the doubly anisotropic gain-media case with material tensors set by $\left[\Lambda\right]= \text{diag}(id/b,-ib/d,-ib/d)$ for $\left|x\right| \leq d/2$. For this particular $\sigma_x=ib/d$ choice, the beamwidth of the Gaussian beam is different from the standard CSP (Fig.~2~(a,e)) and $\mathcal{PT}$ (Fig.~2~(b,f)) cases due to the different imaginary displacement for $(x'_s,y'_s)$. Note that in this case, a Gaussian beam is generated towards both directions (i.e., bidirectionally). In addition, by translating the source inside the slab, one would produce Gaussian beams with different amplitudes on each direction due to the different distances from the source location to the two sides of the slab. Fig.~2(d,h) shows results for the same transformation but now with the point source placed at the left boundary of the slab, $x=-d/2$. In the transformed problem, this is equivalent to an ordinary source for observation points on the same side of slab and to a CSP for observation points on the opposite side of the slab (Fig.~2(d)). Consequently, the Gaussian beam is launched towards the opposite of the slab as a consequence of the field amplification as the wave traverses the slab. Note that the original CSP (i.e., $x_s=ib, y_s=0$) behavior is perfectly reproduced when the source is placed at the boundary of the slab. Thus, both $\mathcal{PT}$ metamaterial and the proposed gain media can mimic CSP perfectly (see Fig~2~(e,f,h) and Fig.~3). This can also be anticipated from the $R'$ profile, where they match perfectly on the right hand side of the slab for Fig~2~(a,b,d). We stress that the comparison made here is between the field produced by a point source placed at the boundary of the metamaterial slab versus the field produced by a CSP in free-space. Based on this, we note that $a_x=0$ choice is crucial in order for the metamaterial slab configuration to mimic the CSP field. Once a phase progression is introduced either by choosing $a_x \neq 0$ (as considered further ahead) or by moving the point source away from the slab, a CSP field cannot be perfectly reproduced. Fig~2~(i) illustrates the basic setup for the FE simulations on a domain truncated by a PML. Fig.~2~(j-l) shows the FE simulation results with metamaterial slabs with constitutive properties set by the respective $\left[\Lambda\right]$.

Fig.~3 shows a more quantitative comparison of the aforementioned cases depicted in Fig.~2(e,f,g,h). Fig.~3(a,b) shows horizontal ($y=0$) and vertical ($x=\lambda+\lambda/100$) (just right of the boundary of the slab) cut of $\Re e[E_z]$ respectively. Note that here we have shifted the position of CSP (Fig.~2(e)) to $x_s=\lambda$ in order to compare the CSP behavior with other cases. As can be seen from both figures, the field behavior of CSP is perfectly reproduced by both $\mathcal{PT}$and gain-only transformation media (Fig.~2(e,f,h)). In addition, by placing the source at the center (see Fig.~2(g)), one obtains a CSP behavior in both directions but now with different imaginary displacements (thus different beam waists). One can also observe from Fig.~3 that although the field behavior on the right side of the slab reproduces the CSP field precisely, the field behavior inside and to the left of the slab is very different in those cases. Although not shown in Fig.~3 for brevity, the agreement between analytical and numerical solutions is excellent.

\subsection{Results for $a_x>0$}
In all the above cases, we have assumed $a_x=0$ which entails material tensors with purely imaginary elements. Next, we show how it is possible to further control the Gaussian-like beam characteristics by varying $a_x$. Fig.~\ref{fig4} shows Gaussian-like beams produced from point sources at various $x_s$ next to doubly anisotropic gain-media slabs based on Eqs.(\ref{to_xyz}), (\ref{Lambda}), and (\ref{tr2}) with different $a_x$. In all cases, $d=\lambda$ and $\sigma_x=-2b/d$.  The plots in Fig.~\ref{fig4}~(a-c) show the respective  $\Re e[E_z]$ distributions based on FE simulations. Fig.~\ref{fig4}~(d) shows the $\Re e[E_z]$ distribution along the horizontal cut $y=0$ for different choices of $x_s$ and $a_x$, namely: $(x_s, a_x) = (-\lambda, 0)$, $(-\lambda,1)$, and $(-1.5\lambda,1)$. Fig.~\ref{fig4}~(e) shows $\Re e[E_z]$  along the vertical cut $x=\lambda/2+\lambda/100$. In Fig.~\ref{fig4}~(e) we also show $\Re e[E_z]$ along the cut at $x=2.5\lambda$ as indicated by the red dotted line in Fig.~\ref{fig4}~(a). Note that for $a_x=0$,  see Fig.~\ref{fig4}~(a) and the blue trace in Fig.~\ref{fig4}~(d), there is no phase progression inside the slab, only amplification. The effect of $a_x$ is to change the electrical thickness and by setting $a_x>0$, a phase progression is produced in the field within the slab, as visible in the  $a_x=1$ case. Note that the $\Re e[E_z]$ distribution along $x=2.5\lambda$  in Fig.~\ref{fig4}~(a) and $x=\lambda/2$ in Fig.~\ref{fig4}~(c) are equal to each other since they correspond to the same amount of amplification and phase progression. Compare also the green and black dotted lines in Fig.~\ref{fig4}~(e). As noted before, by introducing a phase progression (either by changing $a_x$ or by placing the source outside of the slab) the resulting field is not a perfect reproduction of a CSP field anymore. Notice also that in Fig.~4, as the electrical distance between poin source location and the slab boundary is increased the beam waist at the right boundary of the slab is enlarged. Note in particular the field behavior in Fig.~4(e), which shows $\Re e[E_z]$  along the vertical cut $x=\lambda/2$: clearly, as the source is placed further away (or, equivalently, the electrical distance is increased) from the slab, the field on the opposite side of the slab becomes closer to a spherical wave rather than a directed beam. This is due to the fact that slab acts like a launcher and consequently the Gaussian-like behavior is only observable beyond the slab position. This means that as the point source  ('image CSP') is placed further away from the slab, the observed field in the region beyond the opposite side of the slab approaches the far-field of the 'image CSP'. In other words, the field is still associated to the 'image CSP' field, but the observed field (beyond the slab) is the far-field behavior of this source. A similar example was considered in \cite{Savoia_to_16} using a more general coordinate transformation wherein the field on the opposite side of the slab was interpreted as being produced by an `image' CSP. In the present approach, one can also use ($x'_s$,$y'_s$) to define the 'image' CSP that is similar to \cite{Savoia_to_16}. Taken together, these results show that the field due to a point source on the opposite side of the slab can be controlled simply via varying $a_x$ (waist location) and $\sigma_x$ (beamwidth). Finally, we note that by adopting a PML-like CTO transformation (complex coordinate stretching) in Eqs.~1-2, the resulting transformation-media slab is impedance-matched to free-space for all polarizations and incidence angles, regardless of the choices for $a_x$ and $\sigma_x$.

\subsection{Results for $a_x<0$}

Lastly, we consider the choice of negative $a_x$, which entails negative refraction effects~\cite{Pendry_mm_nim_00,Collin_mm_nim_10}. The case $a_x=-1$ and $\sigma_x=0$ in particular recovers the isotropic Veselago slab~\cite{Veselago_mm_nim_68} (other choices of $a_x$ and $\sigma_x$ lead to {\it anisotropic} Veselago slabs~\cite{Kuzuoglu_to_06}) where foci can be established inside and outside the slab. Considering $\sigma_x=-b/d$  as in Eq. (\ref{tr2}), this generalizes to a gain media in which beam waists are created inside and outside of the slab. Fig.~5~(a,d) shows $\Re e[E_z]$  based on the analytical solution from Eqs. (\ref{Ez1}) and (\ref{R}) for metamaterial slabs having $\sigma_x=-b/d$, and with $a_x=-1$ and $a_x=-2$ respectively. Fig.~5~(b,e) shows corresponding FE  simulations results for a point source placed near the associated metamaterial slabs. As a reference, Fig.~5~(c,f) shows the associated analytical results considering $\sigma_x=0$, wherein the two foci, one inside and the other outside the slab are clearly visible. From the plots, we observe that the locations of the Gaussian beam waists in the case $\sigma_x=-b/d$ coincide with the foci location in the $\sigma_x=0$ case with same $a_x$, Moreover, it is seen that this location can be controlled by varying $a_x$. Note also that, although not visible in $\Re e[E_z]$ plots due to the field amplification effects across the slab, in addition to the focus (waist) outside of the slab, there is also one present inside the slab, see the phase profiles insets in Fig.~5~(a,b,d,e). These phase profiles resemble the $\Re e[R']$ of a CSP, which indirectly verifies the mapping of the real point source to a CSP at the focal point of the slab. We should also point out that, for such unusual coordinate mapping, the appropriate solution needs to be built in the transformed coordinates in order to obtain the correct physical solution. When $a_x<0$ the solution will be unphysical if we simply analytically continue the Green's solution as done in Eq.~5. In particular, the solution {\it in the region between the two focal points} needs to chosen based on $H^2_0{\cdot}$ instead of $H^1_0(\cdot)$ in order to satisfy the proper field continuity. This recovers the so-called `source type' branch-cut choice in the CSP literature.

\section{Concluding Remarks}

The contributions from this work are three-fold. First, it was shown that Gaussian beams can be generated from point sources placed inside or outside doubly anisotropic gain-media slabs without the need for $\mathcal{PT}$ symmetry.
Second, it was verified that the location of the equivalent complex point source (CSP), and hence beam properties, can be controlled by varying both the real and imaginary part of the CTO mapping equations. Third, by using negative values for the real part of the CTO mapping equations,it was demonstrated that a real point source placed one side of the metamaterial slab can be mapped to an equivalent CSP on the opposite side of the metamaterial slab. The CSP location is associated to the waist location of the Gaussian beam and can be moved away from the slab. These findings were verified by means of equivalent CSP analytical solutions and by FE simulations employing the derived doubly anisotropic gain-media slabs.

The study here was done in the Fourier domain assuming the linear regime. Because gain media are inherently nonlinear, this means that the present method of analysis is restricted to field amplitudes below the gain saturation threshold. The particular values for the gain-media parameters used here have been chosen for the sake of illustration. Although the resulting gain levels are unrealistic under present technology, more feasible gain levels may be obtained using thicker slabs, as noted in~\cite{Castaldi_to_13}. Nevertheless, the doubly anisotropic gain-media slabs proposed here can be chosen to be homogeneous and hence are inherently simpler than balanced loss/gain $\mathcal{PT}$ slabs, which are necessarily inhomogeneous.


\bibliography{references}

\end{document}